\documentclass[twocolumn]{aa} % for a paper on 2 columns  
\usepackage{graphicx}
\usepackage[authoryear]{natbib}
\bibpunct{(}{)}{;}{a}{}{,}
\usepackage{txfonts}
\begin{document}
\authorrunning{Mar\' \i n-Franch et al.}
\titlerunning{The stellar population of the star forming region G61.48+0.09}
\title{The stellar population of the star forming region G61.48+0.09} 
\author{A.~Mar\' \i n-Franch
          \inst{1}
          %\offprints{A.~Mar\' \i n-Franch}
          \and
A.~Herrero\inst{1,2}
\and
A.~Lenorzer\inst{1}
\and
F.~Najarro\inst{3}
\and
S.~Ramirez\inst{1}
\and
A.~Font-Ribera\inst{1}
\and
D.~Figer\inst{4}
   \institute{Instituto de Astrof\'{\i}sica de Canarias, 38200 La Laguna, Tenerife, Spain\\
              \email{amarin@iac.es}
         \and
	  Departamento de Astrofisica, Universidad de La Laguna, Spain\\
\and
Departamento de Astrofisica Molecular e Infrarroja, IEM, CSIC, Madrid, Spain \\
\and
Rochester Institute of Technology, Rochester, NY, USA}}
 
   \date{Received }

  \abstract
  {We present the results of a near-infrared photometric and spectroscopic study of the star forming region G61.48+0.09.}
  {The purpose of this study is to characterize the stellar content of the cluster and to determine its distance, extinction, age and mass.}
  {The stellar population was studied by using color-magnitude diagrams to select twenty promising cluster members, for which follow up spectroscopy was done. The observed spectra allowed a spectral classification of the stars. }
  {Two stars have emission lines, twelve are G-type stars, and six are late-O or early-B stars.}
  {The cluster's extinction varies from $A_{K_S}$ = 0.9 to $A_{K_S}$ = 2.6, (or $A_{V} \sim$ 8 to $A_{V} \sim$ 23). G61.48+0.09 is a star forming region located at 2.5$\pm$0.4 Kpc. The cluster is younger than 10 Myr and has a minimum stellar mass of 1.5$\pm$0.5$\times$10$^{3}$ M$_\odot$. However, the actual total mass of the cluster remains undetermined, as we cannot see its whole stellar content.}


 \maketitle
%
%________________________________________________________________

\section{Introduction}

The exploration of the Milky Way is experiencing a strong boost due to the use of recent
infrared surveys, like 2MASS \citep{skru06} and GLIMPSE \citep{benjamin03}, frequently combined with data from other 
wavelength ranges, like radio or $\gamma$-rays.  A strong effort is being dedicated to the discovery 
of new stellar clusters, containing massive stars (Arches, Quintuplet, \cite{figer99}, RSGC1, \cite{figer06}), or to a deeper 
study of already known stellar clusters and cluster candidates, resulting sometimes in their identification as massive 
stellar clusters (Cyg OB2, \cite{kno00}, Westerlund1, \cite{clark02}). The search has been strongly enabled by new catalogs 
of stellar cluster candidates like \cite{bica03}, \cite{dutra03}, \cite{mercer05} and \cite{froeb07}. This remarkable effort 
is leading us towards a re-evaluation of 
our ideas about the Milky Way star-forming activity \citep[see f. e][] {figer08}.

\cite{hanson08} review the recent advances, and predict that more clusters are to be 
discovered or identified. Recent examples, like Messineo 1 \citep{messineo08}
and DBS2003-179 \citep{boris08}, seem to confirm this prediction. A difference with former results, however, is that both 
clusters are moderate in their masses (about 2000 and 7000 M$_\odot$, respectively), which indicates that we are discovering representatives
of a probably large number of intermediate mass stellar clusters. This is consistent with what we could expect, as the 
luminosity function of clusters in spiral galaxies follows a declining power law with exponent $\alpha$= -2
\citep{larsen02}, which for clusters of similar age can be translated into the same law for the mass function. These clusters,
being massive and numerous, may play a substantial role in the star formation history of the Milky Way and other galaxies.

This is the first paper of our MASGOMAS (MAssive Stars in Galactic Obscured MAssive clusterS) 
project, a study of massive star cluster candidates in the Milky Way that is being carried out using LIRIS at WHT.
Our idea is to select known clusters or cluster candidates from the literature that could turn out to be massive
clusters, observe them photometrically and perform a spectroscopic follow-up of the most promising ones.
In this paper we present a photometric and spectroscopic study of the stellar population of 
G61.48+0.09, a cluster that has got recent attention \cite{puga04} and whose distance is matter of debate (see below).

\begin{figure*}
\centering
\includegraphics[width=15cm]{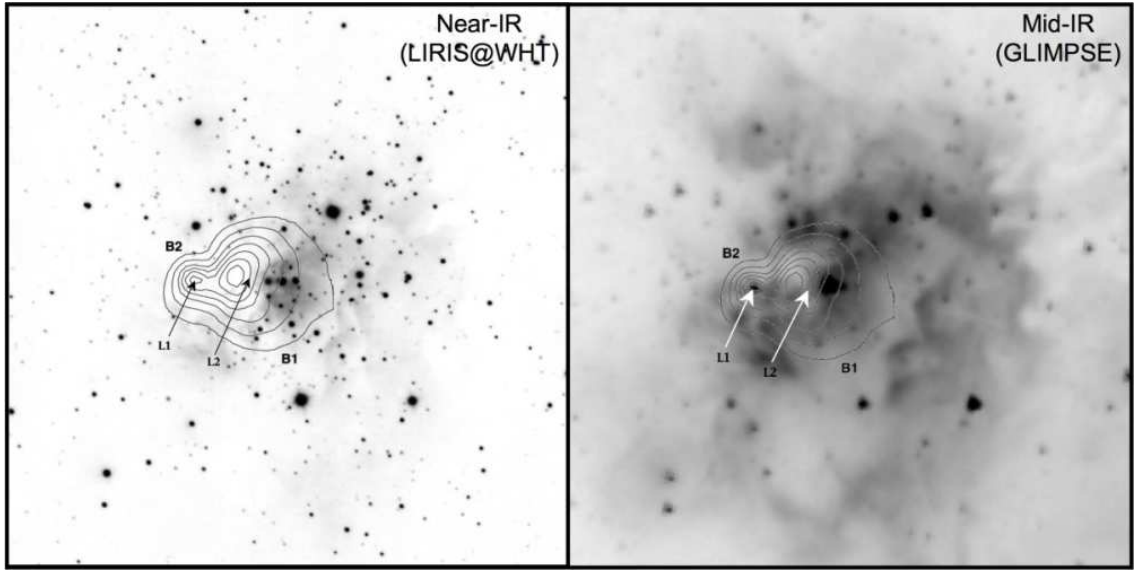}
   \caption{Near$-$ ($K_S$,  LIRIS@WHT) and mid$-$ (3.6 $\mu$m, GLIMPSE) infrared images of the star forming region G61.48+0.09. The grey contours correspond to the \cite{garay98} continuum emission at 8.3 GHz. The positions of stars L1 and L2 is also marked.}
      \label{introID66}
\end{figure*}

G61.48+0.09 is a complex star$-$forming region located in the direction
of the Cygnus arm. Figure~\ref{introID66} shows a near$-$ (left panel) and a mid$-$ (right panel) infrared images of the G61.48+0.09
region. Continuum emission contour levels at 8.3 GHz \citep{garay98}
are also shown. As it can be seen in the right panel of this figure, the G61.48+0.09 region consists of some bright stars surrounded by a HII region, Sh2-88, composed of two sub-regions, B1 (westwards) and B2 \citep[eastwards; ][]{deha00}. 
The central part is a triplet of stars. Although the region has been extensively studied by different authors, no spectroscopy of their stars exists. \cite{puga04}, in their polarimetric study of the region,present a detailed review of previous radio and near$-$IR photometric work. From their polarimetric study and the energetics balance they conclude 
\citep[as did][]{deha00} that the main ionizing source of this region is probably still hidden
behind obscuring clouds of natal material. A strong source in the L' band, identified as L1, and a fainter L-source identified as L2, are proposed as possible main ionizing sources. L1 and L2 locations have also been marked in Figure~\ref{introID66}. In the right panel, L1 can be clearly seen as a bright object in the center of the HII subregion Sh2-88B2.

The distance to G61.48+0.09 is somewhat controversial. The first determination was 
a kinematic distance derived by \cite{reif70}, who using the \cite{schmidt65} model of the Galaxy arrived at 
the two possible distances of 2.0 and 7.5 Kpc. Since then, many authors have made new determinations
obtaining different values, but always in rough agreement with one of both values obtained
by Reifenstein. \cite{churchwell90} give a distance of 5.4 kpc based on ammonia observations 
combined with the identification of the region with CO observations from \cite{solomon87}, who use 
different techniques to discriminate close and far distances in the list of giant molecular 
clouds from \cite{sanders86}. A comparison of different determinations may be found in \cite{deha00},
that adopt the short distance of 2.5 kpc, as most authors do. Nevertheless, the distance from \cite{churchwell90}
has very recently being adopted by \cite{zhu08} in their study of Galactic clusters. Of course, many derived global
properties of G61.48+0.09, like its total stellar mass or energetic content, may strongly depend
on the adopted distance.

 We present
new NIR photometric data and, for the first time, NIR spectroscopy of stars in the cluster
region. Section~\ref{obs} presents the observations and data reduction. In Section~\ref{results} we present 
the results of the photometric (\ref{cmd}) and spectroscopic (\ref{specres}) analysis. 
Section~\ref{discus} presents the discussion of these results and in Section~\ref{conc}
we present our conclusions.

\section{Observations and data reduction}
\label{obs}

This study is based on broad-band imaging, long-slit and multi object spectroscopic observations done with LIRIS, a near-IR imager/spectrograph mounted at the Cassegrain focus of the 4.2m William Herschel Telescope (La Palma). Table~\ref{TableObs} gives a summary of the observation details.

\subsection{Photometry}

\begin{figure*}
\centering
\includegraphics[width=15cm]{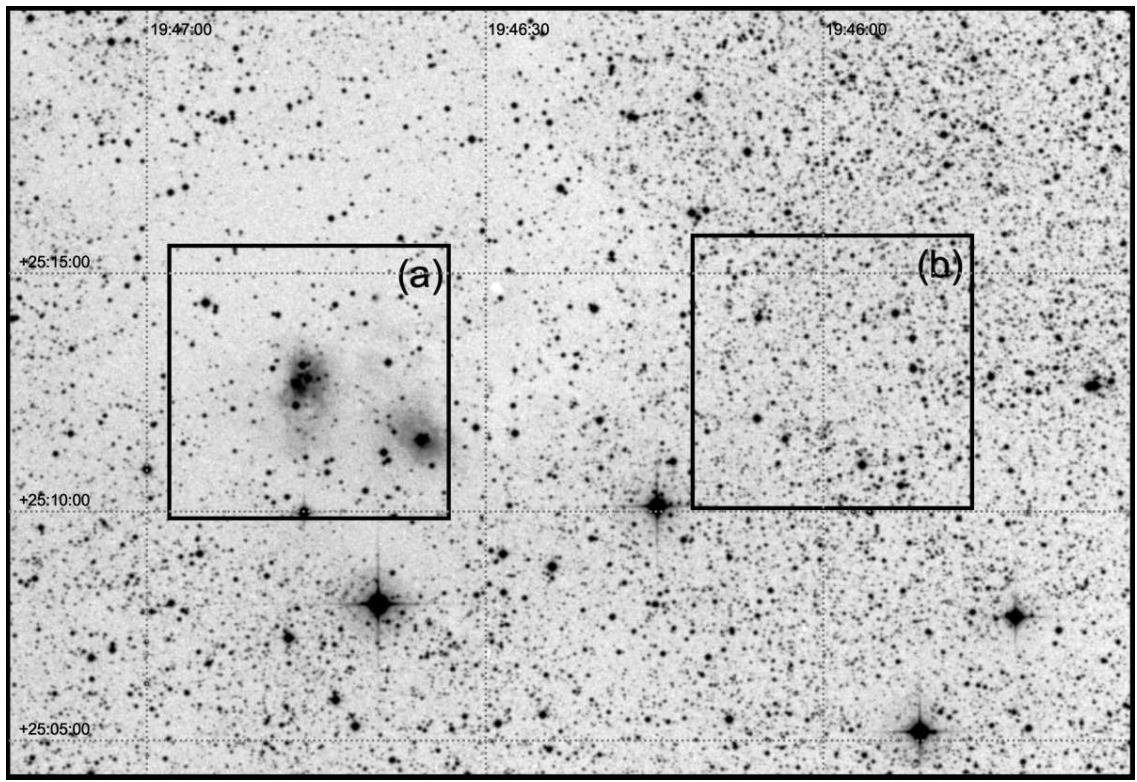}
   \caption{Wide field DSS-2-infrared survey image of the G61.48+0.09 star forming region. The squares represent the observed target (a) and control (b) fields of view.}
      \label{FoVs}
\end{figure*}

LIRIS is equipped with a Hawaii 1024 $\times$ 1024 HgCdTe array detector, with a spatial scale of 0.25 $\arcsec pixel^{-1}$, providing a field of view of 4.27 $\arcmin$ on a side. Broad-band imaging observations were performed using the $J$, $H$ and $K_S$ filters. In order to optimize cosmic rays/bad pixels rejection and sky computation, a standard dithering mode (8-point pattern) was used. Two different fields have been observed: the target field, centered in the G61.48+0.09 region, and a control field located $\sim$8$\arcmin$ west from the target field. Figure~\ref{FoVs} shows a DSS-2-infrared survey image in which both target (a) and control (b) fields have been marked. It can be seen that there is a lack of stars around the G61.48+0.09 field if compared to the control filed. This suggests that there is a cloud associated to the G61.48+0.09 region which is responsible for the increased extinction of the background field stars. The data were reduced using the FATBOY \citep{Eiken06, Warner09} and LIRIS-QL image reduction packages following standard NIR reduction procedures. The average seeing during the imaging observations of the G61.48+0.09 region was $\sim$0.6$\arcsec$. In the left panel of Figure~\ref{introID66}, the resultant G61.48+0.09 image in the $K_S-$band is shown.

\begin{figure}
\centering
\includegraphics[width=15cm]{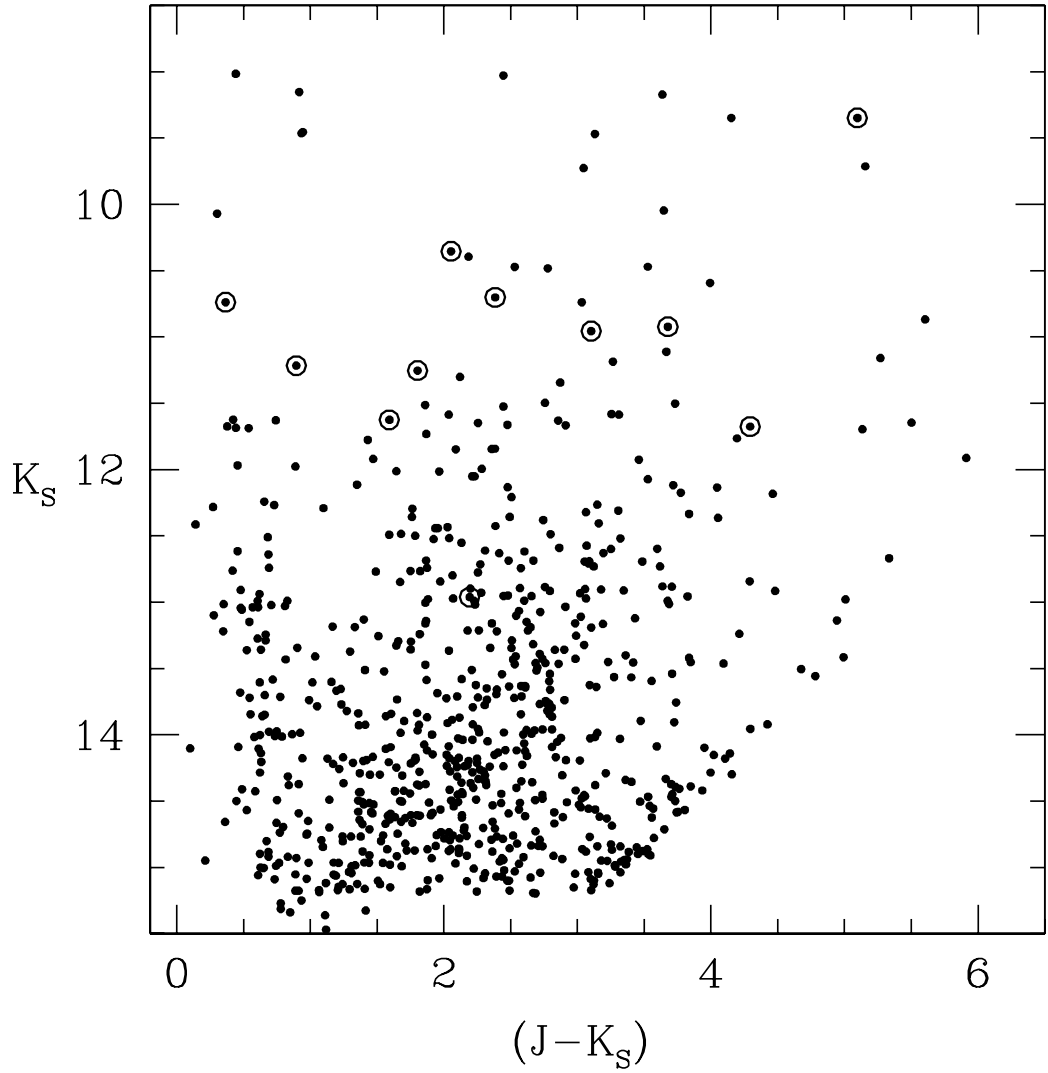}
   \caption{2MASS CMD of the target field. The stars used for the photometric calibration are marked with open circles.}
    \label{starSelection}
\end{figure}

\begin{figure}
\centering
\includegraphics[width=15cm]{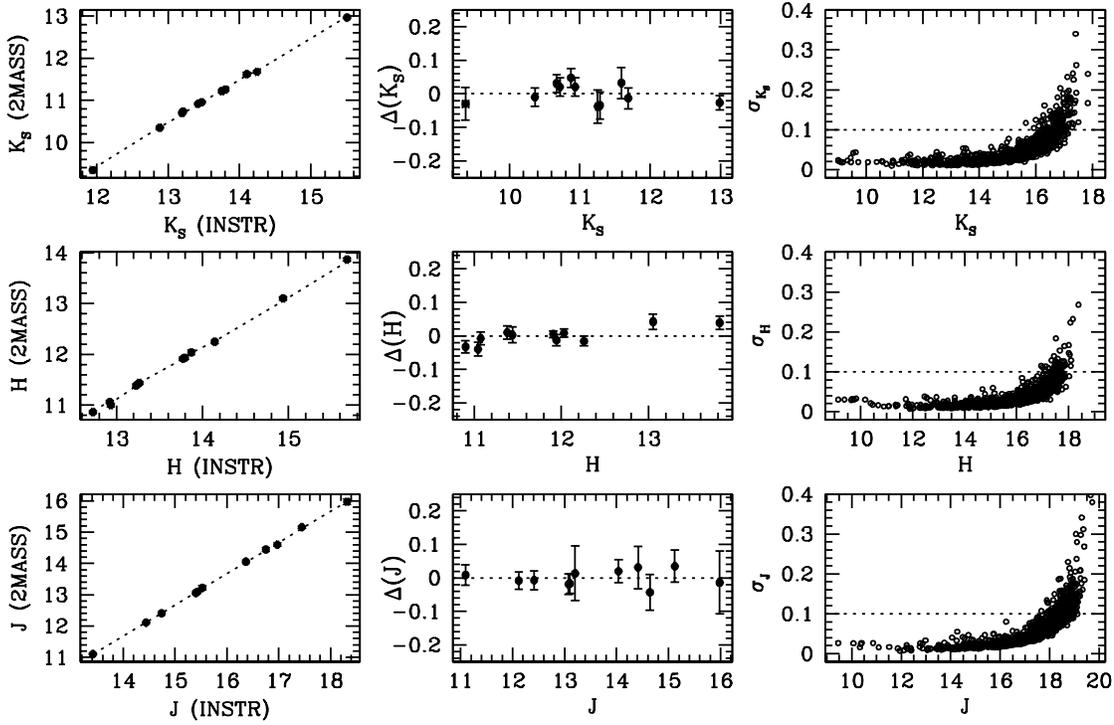}
   \caption{Photometric calibration in the $J-$, $H-$ and $K_S-$bands for the G61.48+0.09 field. Left panels show the photometric calibration from instrumental (INSTR) to the calibrated magnitudes (2MASS).  Central panels show the difference between the calibrated photometry and the 2MASS photometry. Right panels show the PSF fitting $\sigma$ parameter provided by ALLFRAME.}
    \label{cal066}
\end{figure}

DAOPHOT II, ALLSTAR, and ALLFRAME \citep{Stetson94} were then used to obtain the instrumental photometry of the stars. The photometric lists of the observed targets were cleaned of non-stellar objects and poorly measured stars on the basis of the SHARP and PSF fitting $\sigma$ parameters provided by ALLFRAME. Only stars with $|$SHARP$| < 0.25$ and $\sigma$ $<$ 0.1 have been considered. The photometric calibration was based on the 2MASS photometry. Figure \ref{starSelection} shows the 2MASS color$-$magnitude diagram (CMD) of the target field. For the calibration, a set of isolated, non-saturated stars for which good photometry exists in both 2MASS and our images was selected.  These stars have been marked in Figure \ref{starSelection} with open circles. They span a wide range of magnitudes ($13 > K_S > 9$ mag) and colors ($0 < (J-K_S) < 5$). Comparing their instrumental photometry with the 2MASS catalogue, a calibration equation (which includes a color term) was derived for the $J$, $H$ and $K_S$ filters. Figure \ref{cal066} shows the photometric calibration results for our target field. Left panels show the photometric transformation from instrumental (INSTR) to calibrated 2MASS magnitudes.  Central panels show the difference between the calibrated photometry and the 2MASS photometry. It can be seen that this difference is smaller than 0.05 mag for most of the selected stars. Right panels show the PSF fitting $\sigma$ parameter provided by ALLFRAME. This calibration procedure has been repeated for the control field. Finally, as the few stars brighter than $K_S\sim10$ appear saturated in our data, their magnitudes have been obtained directly from the 2MASS catalogue. 

It is worth mentioning that the 2MASS photometry for most of the stars brighter than $K_S \sim 11$ and redder than $(J-K_S) \sim 3$ is strongly affected by the molecular cloud. For those stars, the 2MASS catalogue is adding the emission of dense cloud clumps to the star photometry, making the star artificially brighter and redder. In fact, some of the 2MASS detections in this region of the CMD are not real stars, but molecular cloud clumps instead. This effect is due to the limited 2MASS spatial resolution. For this reason, it is not possible to increase the number of stars with $(J-K_S)> \sim 4-5$ used for the photometric calibration. The calibrated photometry is then less reliable for stars with $(J-K_S) > 5$. But, as it will be shown in Sections~\ref{results} and \ref{discus}, these stars are background late type stars that have not been used for the cluster description, with the exception of a YSO. As a conclusion, this lack of highly reddened stars in the photometric calibration is not affecting cluster results significantly.

\subsection{Spectroscopy}

\begin{figure}[hb!]
\centering
\includegraphics[width=12cm]{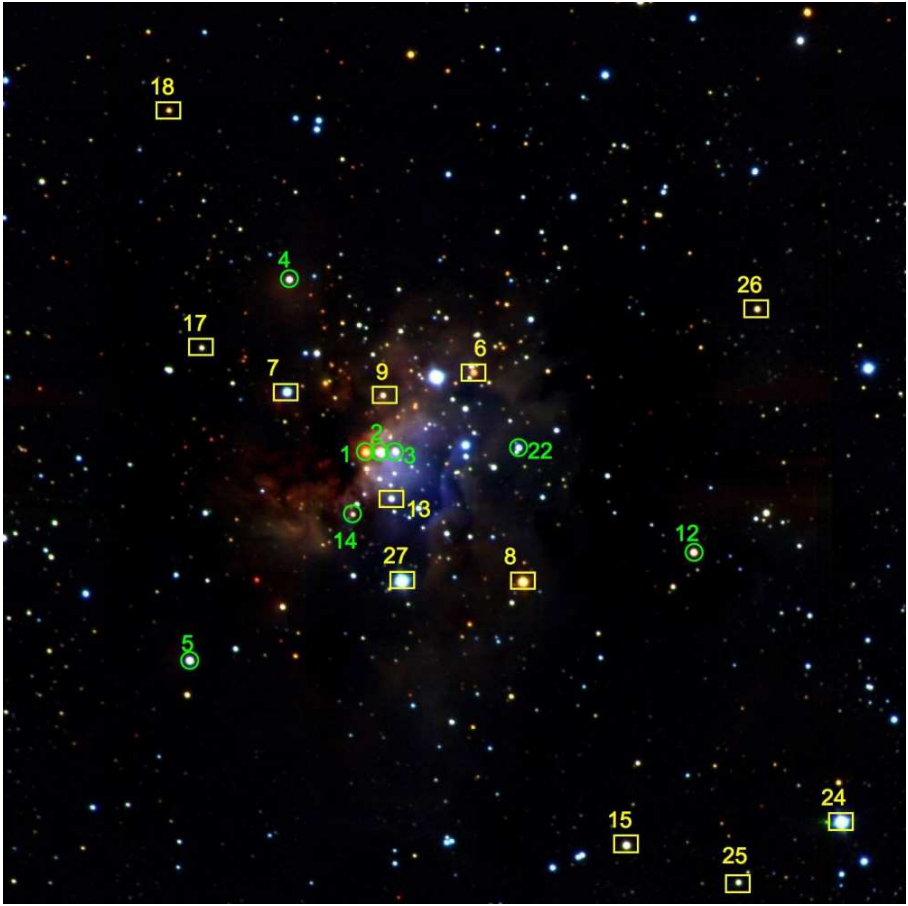}
   \caption{False color composition image of the central part of the G61.48+0.09 field. The position of the spectroscopically observed stars si shown. The early-type stars (see Table~\ref{spectab}) have been marked with circles, and the rest with squares}
      \label{MOSim}
\end{figure}

We obtained medium$-$resolution $HK-$ and $K-$band spectra with WHT/LIRIS in multi$-$object (MOS) and long-slit (LS) spectroscopic modes. The slit width was 0.8$\arcsec$, allowing a spectral resolution of $\lambda / \Delta \lambda \sim $ 945 and 2500 in the $HK-$ and $K-$band spectra, respectively. We observed in a standard nodding pattern ABBA with individual exposure times of 20, 40 and 120 s ($HK$) and 120 s ($K$). Table \ref{TableObs} lists the details of the spectroscopic observations, and Fig.~\ref{MOSim} shows a false color composition image of the G61.48+0.09 field marked with the position of the spectroscopically observed stars (see Section~\ref{specres}). 

The observations were reduced as follows: We first scaled B fields to their corresponding A fields using the quotient of medians as scaling factors. This step crudely corrects for average sky variations between frames with large integration times. We then subtracted B frames from A frames, and A frames from B frames before applying flatfield correction. 
The AB and BA frames were shifted in the spatial direction so that all positive continuum traces coincide before all frames were combined into a master frame. We then extracted the spectra from the master frame. The spectra were correlated with that of the standard star yielding wavelength solutions accounting for distortions of high orders. This solution was then applied to the spectra before division by the spectrum of the standard star from which the stellar features had been removed.

\section{Results}
\label{results}

\subsection{Color-Magnitude diagrams}
\label{cmd}

\begin{figure}
\centering
\includegraphics[width=9cm]{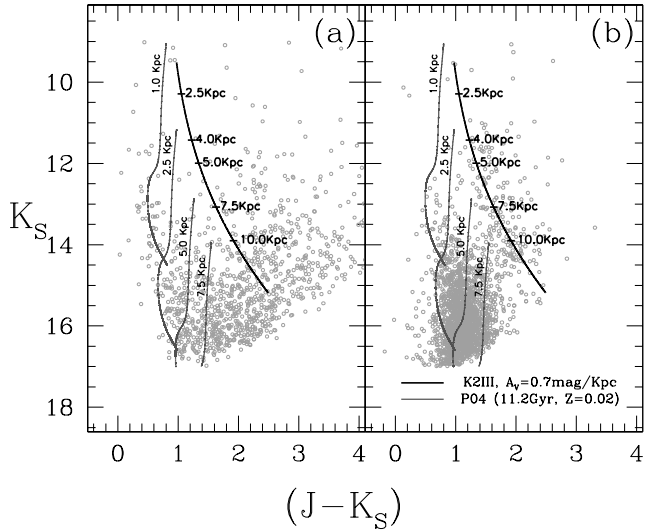}
   \caption{Color-Magnitude Diagrams of the target (a) and control (b) fields. The black solid line shows the expected position of a K2III star as a function of its distance. To construct the K2III sequence, data for $M_V$ have been taken from \cite{cox00} and IR intrinsic colors from \cite{ducati01}. Grey lines represent a fraction of a 11.2 Gyr and solar metallicity isochrone, located at different distances.  The \cite{pie04} stellar evolution library and the \cite{CK03} bolometric correction library have been used to generate the isochrones. Besides, the \cite{rieke89} extinction law and R=3.09 \citep{rl85} have been assumed. }
   \label{cmds}
\end{figure}

The CMDs of the target and control fields can provide valuable information about the G61.48+0.09 region's extinction, distance and stellar content.  Figure~\ref{cmds} shows the CMDs corresponding to the G61.48+0.09 region (a) and the control field (b). 

The position of a star on the CMD is determined from its absolute magnitude, intrinsic color, distance and extinction. When comparing target and control fields CMDs, the first major difference that can be observed is the absence of the red-clump strip in the target field, while this feature can be clearly seen in the control one. The red-clump strip is the CMD trace that is populated by red-clump stars located at different distances. \cite{lopc02} developed a method to use red-clump stars as a distance indicator. In essence, this method uses a K2III star as representative red-clump star and, assuming an extinction law, computes the expected position of this star in the CMD when located at different distances. Comparing the former with the observed red-clump strip, the distance of red-clump stars can be estimated. 

In this paper we adopt the \citet{cox00} absolute $V$-magnitude and \cite{ducati01} IR colors for a K2III star. The black solid line in Figure~\ref{cmds} shows the expected position of a K2III star as a function of its distance. The \cite{rieke89} extinction law and R=3.09 \citep{rl85} have been assumed. The fact that the red-clump strip is not seen in the G61.48+0.09 region shows that the cluster, in particular its associated cloud, is adding extinction to the background red-clump stars. This indicates that the cluster is located between these red-clump stars and the Sun. In other words, an upper limit to the distance can be established by comparing the target and control CMDs. As a result, it can be concluded that the G61.48+0.09 region is located at a distance less than $\sim$4Kpc. It is worth mentioning that this rough estimation depends on the assumed intrinsic colors and magnitudes of a K2III star, as well as on the assumed extinction law. 

On the other hand, the local main sequence (MS) can be clearly seen in both CMDs, although for magnitudes fainter than $K_S \sim $14, it appears more affected by extinction in the target field than in the control one. This additional extinction is due to the presence of the star forming region G61.48+0.09.  At this point, a simplistic approximation can be done. \cite{binney00} find the age of the solar neighborhood to be 11.2$\pm$0.75 Gyr. If we assume that  the solar neighborhood is composed mainly of stars with this age and solar metallicity, then a suitable isochrone can be over-plotted in Figure~\ref{cmds}. In this context, a portion of a 11.2 Gyr and solar metallicity isochrone, located at different distances has been over-plotted.  The IAC$-$star \citep{aparicio04} synthetic CMD program has been used to compute the isochrone. \cite{pie04} stellar evolution library and \cite{CK03} bolometric correction library have been considered together with the \cite{rieke89} extinction law. Comparing these isochrones with the CMDs, it can be observed that the main difference between the target and control MSs starts at a distance of $\sim$2.5 Kpc. That is, local MS stars located at a distance shorter than $\sim$2.5 Kpc are not affected by the presence of the star forming region G61.48+0.09, but more distant MS stars have an increased extinction when compared to the control field. 

As a result, this simple argument leads to a distance estimation of $\sim$2.5 Kpc for the star forming region G61.48+0.09. In Section~\ref{discus}, a more precise distance determination of G61.48+0.09 is derived based on the spectroscopic analysis combined with photometric information, but it is interesting how this photometric analysis is able to provide a first approximation. Kinematical studies are compatible with distances of 2.0 and 6.5 Kpc, so this photometric analysis points towards G61.48+0.09 located at a distance closer to the smaller value.  

It is worth mentioning that no trace of a cluster MS is found in the G61.48+0.09 region's CMD. As it will be shown later, this cluster is affected by a high differential reddening, which explains the absence of a cluster MS. This differential reddening is shifting every cluster member in a different way along the reddening vector, so the cluster MS is highly blurred.
 
\subsection{Spectral classification}
\label{specres}

The G61.48+0.09 region's CMD has been used to select promising cluster members, for which follow up spectroscopy has been done. The selection was done based on the location of these stars in the CMD and their position in the cluster. In other words, the LS and MOS candidate list was performed based on photometric and spatial information. 
Note that this selection depends actually on the field morphology, as it was mainly 
observed with multiobject spectroscopy.

\begin{figure*}
   \centering
   \includegraphics[width=16cm]{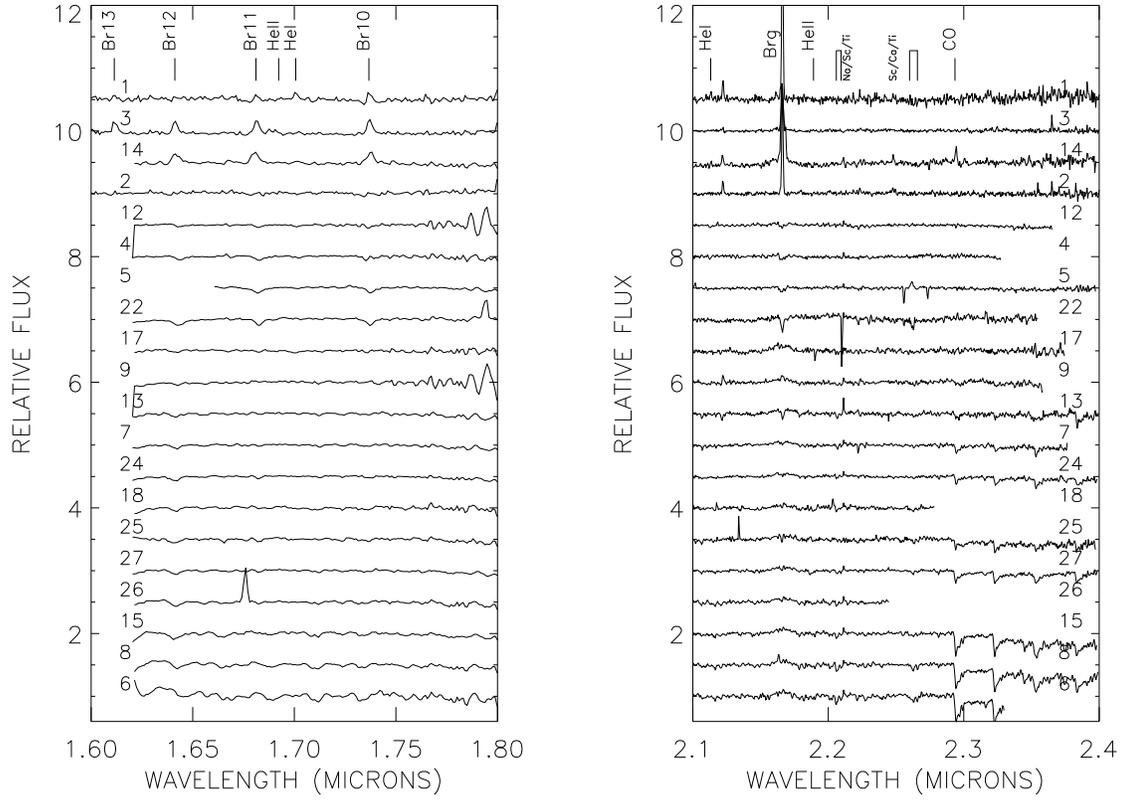}
      \caption{H and K spectra of observed stars in the G61.48+0.09 star forming region.}
         \label{specfig}
   \end{figure*}

The observed spectra are presented in Figure~\ref{specfig}. Two objects are classified as emission
line objects, twelve turned out to be red objects, mainly G-type stars, and six resulted to be OB stars,
from O9 to B3-B5. For the classification, the catalogs of \cite{hanson96}, \cite{hanson98}, 
\cite{wallace97} and the relation between spectral type and equivalent width of the CO band head
given in \cite{davies07} were used. 
The more recent atlas of \cite{hanson05} or the high resolution atlas of \cite{wallace96}
were not used, as their resolutions (R$\sim$8000) are much higher than ours .
 
\begin{figure}
   \centering
   \includegraphics[width=8cm]{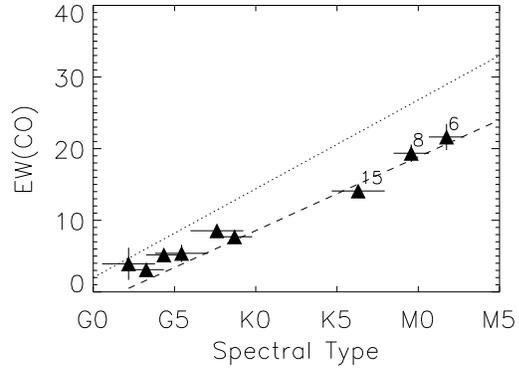}
      \caption{EW of the CO band head versus spectral type. Dotted line is the relation for supergiants presented
      by \cite{davies07} and the dashed line is the corresponding relation for giants by the same authors. }
         \label{EW_CO}
   \end{figure}

Cool stars were first classified in spectral type using the \cite{wallace97} catalog. 
To this aim, the CO band and the features around 2.265 $\mu$m (Ca/Sc/Ti), 2.207 (Na/Sc/V)
and 2.09 (CN) were used as primary indicators. Typical uncertainties (estimated from
independent determinations by two of us) are $\pm$1 spectral subclass. After determining the spectral
type, we determined the luminosity class using the Spectral Type-EW(CO) relation in \cite{davies07}.
This is illustrated in Figure~\ref{EW_CO} where we show the EW of the CO band head in our stars and Davies
relationship. The CO EW was determined in the same way as indicated in \cite{davies07}: we measured the
EW between 2.294 and 2.304 $\mu$m and adopted the average level between 2.288 and 2.293 $\mu$m
for the continuum. Errors were estimated from the local continuum rms.

We see that the K and M stars fit clearly the giant calibration. The situation is less clear
for the G stars, but their positions in the CMD indicate that they are not brighter than the blue
stars in the sample. Therefore, they cannot be red-yellow supergiants. This same argument can be 
applied to stars $\#18$ and $\#26$, for which we could not get
the spectra of the CO band (because of the usual displacement of the spectra when the slits
are displaced from the mask central line), and for star $\#17$ (F-type), where the CO-band
is not visible. Additionally, we note that from the observed spectrum, 
star $\#17$ could be a main sequence object, but then it would be too bright when corrected 
from reddening (except if it has a particularly strong differential obscuration). The spectra, magnitudes
and CO band head equivalent widths are thus compatible with all stars of type later than A, being
yellow or red giants. 

Early type spectra were classified using the \cite{hanson96} and \cite{hanson98} atlases. The main indicator used
was the presence of HeI lines at 1.701 and 2.113. They are present in all stars we
classified as early-type stars, except in star $\#22$. According to \cite{hanson96} and \cite{hanson98}
these lines begin to disappear around B3/B5 in dwarfs. For supergiants the HeI 1.70 line is present for all
spectral types and the HeI 2.113 dissappears around B8/B9. 
The behavior of the Br series as compared with the other early-type
stars points to a classification as B3/B5, probably of luminosity class V. Stars $\#4$, $\#5$ and
$\#22$ were thus classified as B1, B2 and B3-B5 respectively. Star $\#12$ shows in addition
a trace of HeII 2.189 and was classified as O9.

We tried to derive the luminosity class for the early types directly from their spectra, but the low resolution
in the H$-$band and the uncertainties in the reduction of the Br$_\gamma$ line in the K band prevented
us from using the broadening of the H lines for the classification. We also tried to use the ratio
of the Br11/HeI 1.701$\mu$m. For star $\#12$ this is difficult, as we are in the region of 
lower sensitivity for this indicator. Stars $\#4$ and $\#5$ have values favoring luminosity
class V, while star $\#22$ is clearly of luminosity class V according to this ratio.

However, we can still use an indirect argument to assign a luminosity class to star $\#12$.
The stars $\#8$ and $\#6$ are red giants and they have very large extinctions. This
indicates that they are behind the group of early-type stars. Therefore, they limit the possible distance
to star $\#12$ and, consequently, its absolute magnitude, which correspond then to a
luminosity class V stars. The same argument can be applied to the rest of the stars in the cluster.
Moreover, this conclusion is consistent with our previous analysis of the CMD
(see Sect.~\ref{cmd}). If the early type stars were supergiants, then the derived cluster distance 
would be of the order of 7-8 kpc. This is not consistent with Fig.~\ref{cmds}, where the 
CMD of the target field does not show the red clump strip for distances between 3 and 7 kpc.

If the early-type stars were just giants the result would still be inconsistent.
Star $\#22$, and most probably also stars $\#4$ and $\#5$, are dwarfs. Assuming the
stars $\#12$ and $\#2$ (that is classified as B0V below) to be giants would result in two different distances for the
two groups. Stars $\#22$, $\#4$ and $\#5$ are located at a distance of $\sim$ 2.5 kpc, while
stars $\#12$ and $\#2$ would be at 3.5 kpc. This argument leads us to a luminosity class V for 
all the early-type stars.

The spectra of stars $\#1$, $\#2$, $\#3$ and $\#14$ show strong emission lines. The first three are in
the central part of the cluster, and have been observed using a long slit that included
all three stars. Unfortunately, the nebular emission in that region is very strong and highly
spatially variable, so that at our low resolution their spectra are contaminated by emission lines, particularly
in Br$_\gamma$. A similar problem arises with star $\#14$, that was observed in the MOS mask
and lies in the B1 subregion. We cannot separate the nebular and stellar lines.
 
Star $\#3$ displays a spectrum that resembles that of star $\#22$ (i.e., without any
clear He features) except for the emission in the Bracket series. We classify it as a B3-B5 star 
contaminated by the strong nebular emission.

Stars $\#2$ shows a weak absorption at HeI 1.701 and a weak Br11 line, without trace of HeII
absorption, which at this resolution is consistent with an O9 or B0 spectral type. This is also consistent
with the K-band spectrum showing no absorption features, with a moderate emission in
Br$_\gamma$ and HeI 2.059, and with the analysis of \cite{puga04} that estimate a B0V
spectral type for this star. In our CMD this star falls between star $\#12$ and $\#4$ that we classify
as O9V and B1V. Therefore we classify star $\#2$ as B0V.

Star $\#1$ (star $\#82$ of \cite{deha00} and \cite{puga04}) 
is the outstanding red object at the center of the cluster and it could be its main ionization
source. In such case, its spectrum would correspond to that of an early O dwarf. However, we
find no trace of HeII lines or CIII/NIII/OIII emission at 2.116, indicating that the spectral type is later than O7.5.
At the same time, we see no CO absorption, so the star has to be of mid-F type or earlier. If the 
extinction law towards the object is normal, then the spectrum and the observed
K magnitude would be compatible with an early F supergiant, but then the star should be 
much redder in the color-color diagram. Therefore we conclude that the object
has a spectral type between early-A and late-O. We cannot make a better classification, but we note that the 
K-band spectrum resembles that of the YSO object in G118.796+1.030 presented by \cite{bik05}, so we classify it as an YSO. 
This is consistent with the strong reddening present in this star, and with the difficulties pointed out
by Puga et al. to fit its SED. This classification, however, indicates that this is not the main ionization
source of the cluster, that is probably one of the hidden sources \citep[see][for a more detailed
discussion]{puga04}.

Star $\#14$ is quite similar to star $\#1$ in 
the K-band, but with a stronger and broader emission in the Br$_\gamma$ line, as well as in 
the Bracket series in the H-band. We classify it as Be as we cannot identify further features.
Stars $\#1$ and $\#14$ have very large (J-Ks) colors. For star $\#14$ this is not a problem, because
it still falls within our calibration stars range. Star $\#1$ on the contrary falls beyond this limit, but we note that
star $\#6$, which falls in the same region, has infrared colors consistent with its spectral classification
and a very large reddening. We thus think that the colors of star $\#1$ are reliable. The color excesses
are consistent with disks around star $\#14$ \citep{gehrz74} and star $\#1$ \citep{jiang08}.

The derived spectral types and some cross$-$correlations with previous works can be found in Table~\ref{spectab}. 

\section{Discussion}
\label{discus}

The spectral classification of the early OB stars found, together with their photometry, allows the determination of
various cluster parameters, such as its extinction, distance, age and mass. 

\subsection{Extinction, distance and age}

Once the spectral type and luminosity class of a star is derived, its intrinsic color and absolute magnitude is known. Comparing them with its apparent color and magnitude, the extinction of this star can be derived. Assuming the \cite{rieke89} extinction law with R=3.09 \citep{rl85}, the extinction, $A_{K_S}$, of a star can be determined using the following equations:

\begin{equation} \label{eq1}
A_{K_S} = {{E_{J-K_S}} \over {1.514}} = {{E_{H-K_S}} \over {0.561}}
\end{equation}

\begin{figure}[ht!]
\centering
\includegraphics[width=8cm]{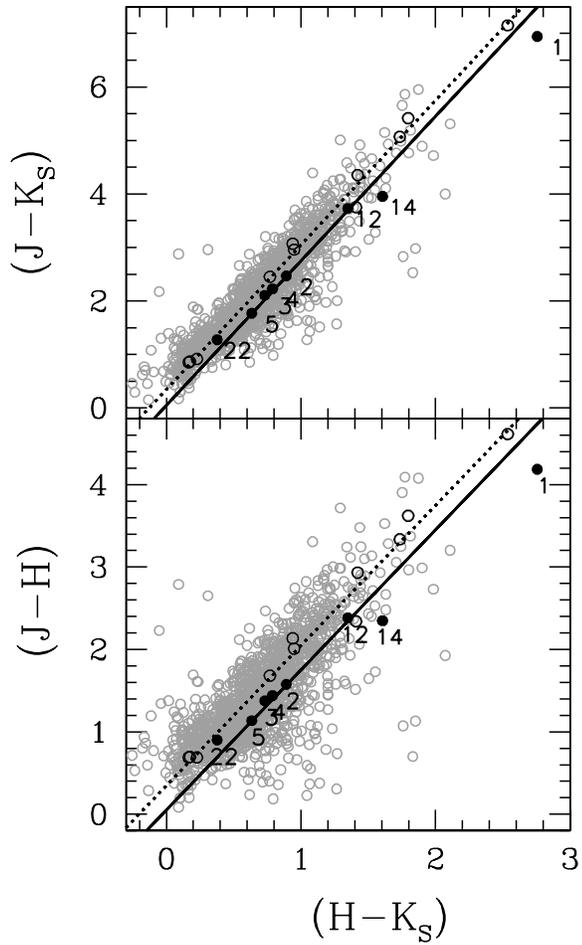}
   \caption{Color-color diagrams of the G61.48+0.09 region (target field). Open and filled circles represents the late and early type stars, respectively, listed in Table~\ref{spectab}. The solid line
   indicates the direction of the reddening vector for a B2V star, while the dotted lines corresponds to a G6III. \cite{ducati01} IR intrinsic colors and the \cite{rieke89} extinction law have been adopted.}
      \label{colorcolor}
\end{figure}

As the intrinsic colors of the cluster stars are known from their spectral classification, $A_{K_S}$ can be derived from the color$-$color diagram. Figure \ref{colorcolor} shows the color-color diagrams for the stars in the G61.48+0.09 target region. The solid and dotted lines show the reddening vector from the expected intrinsic colors of a B2V and G9III star, respectively, assuming the extinction law of \cite{rieke89}. Filled and open circles correspond to the stars classified in Section~\ref{specres} as early (O9V-B5V) and late (G3III-M3III) type stars, respectively. It can be seen how the assumed extinction law reproduces very well the slope in the color$-$color diagram for both the early and late type stars. This suggest that the extinction law assumed for the dense molecular cloud is not very different from that in the interstellar medium. We also see that the early-type stars are affected by a strong differential reddening. The most reddened early-type star ($\#$12) is actually far from what we would identify as the center of the cluster (centered at star $\#$2). This is an indication that the cluster is still sweeping out the original molecular material. Besides, stars  $\#$1(YSO) and  $\#$14 (Be) present a high infrared excess. 

We adopt the \citet{cox00} absolute $V$-magnitude and \cite{ducati01} IR colors for the early type stars except for star $\#12$ (O9V), for which \citet{martins05} observational scale colors have been adopted. The value of $A_{K_S}$ has been obtained for each star using equation~\ref{eq1}, and the mean of the two results is considered as the final $A_{K_S}$. Table~\ref{ageTab} lists $A_{K_S}$ results in column 1. It can be seen that the analyzed stars' extinctions vary from $A_{K_S}$ = 0.9, or $A_{V} \sim$ 8, to $A_{K_S}$ = 2.6, or $A_{V} \sim$ 23. That is, the extinction is highly variable across the G61.48+0.09 star forming region. 
 
Once the extinction of the individual stars is known, their distances can be derived comparing their apparent and absolute magnitudes. Column 2 in Table~\ref{ageTab} lists the derived distances for the six cluster stars. The mean obtained distance is 2.5 Kpc. The uncertainty associated to the distance determination has several contributions. On one side, any uncertainty in the photometry of the stars, in their assumed intrinsic colors and absolute magnitudes have an effect on the distance result. On the other side, the dominant source of error comes from the extinction. As it has been mentioned before, the \cite{rieke89} extinction law with R=3.09 \citep{rl85} has been adopted. The interstellar extinction law could be different in dense molecular clouds. For example, \citet{BGH81} found an anomalous extinction law in Orion, and \citet{Chi81} in the Ophiuchus dark cloud. So the interstellar extinction law for the cluster stars could be different from the adopted one, although not much according to Figure~\ref{colorcolor}. If that is the case, any differences between the adopted and real extinction laws would affect the derived distance. For this reason, we consider the sigma of these individual distances as the final uncertainty, that is, we adopt 2.5 $\pm$ 0.4 Kpc as the distance of the cluster. This distance, 2.5 Kpc, is fully compatible with the shorter distance derived in previous kinematical studies and with the analysis of the CMD from previous section. 

\begin{figure*}[ht!]
\centering
\includegraphics[width=12cm]{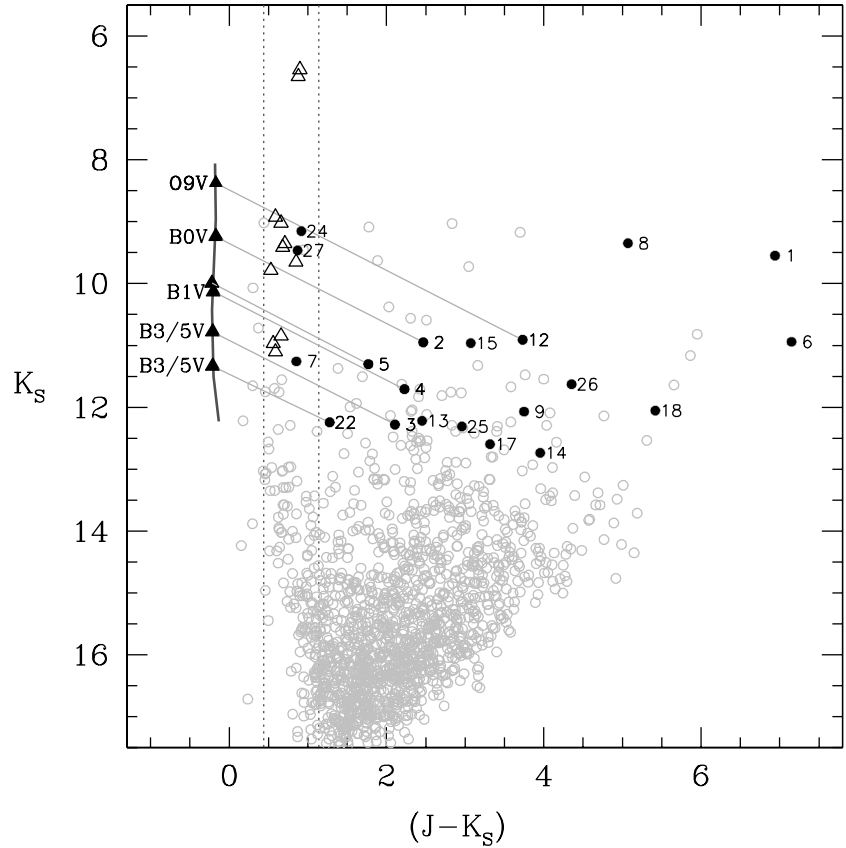}
   \caption{Calibrated color$-$magnitude diagram of the G61.48+0.09 star forming region. The spectroscopically observed stars have been marked with filled circles.  Filled triangles show the position of the reddening corrected early type stars, that is, they show the reddening corrected cluster main sequence. Open triangles show the reddening corrected position of the late type stars. The solid line shows the zero$-$age main sequence located at a distance of 2.5 Kpc. Vertical dashed lines show the intrinsic color interval of the late type stars with spectral types between G0III and M3III.}
   \label{id066CMD}
\end{figure*}

Figure \ref{id066CMD} shows the calibrated CMD obtained for the star forming region G61.48+0.09.
The spectroscopically observed stars are marked with filled circles (the numbers correspond to our own listing).
In this CMD we can see the local main sequence formed by slightly reddened field stars while cluster stars are expected
to have larger reddening. This figure also shows the reddening corrected
position of the spectroscopically observed stars. Filled triangles show the position of the reddening corrected 
early type stars, that is, they show the reddening corrected cluster MS. Solid lines show the zero$-$age MS 
located at a distance of 2.5 Kpc. Open triangles, on the other side, show the reddening corrected position of the late type stars.
Vertical dashed lines show the color interval of the late type stars with spectral types between G0III and M3III.

Regarding cluster's age, only an upper limit can be established with present data. The earliest star found in this cluster is a O9V. If we assume that all stars in the cluster were formed at the same time, and we adopt the calibration of stellar parameters given by \cite{martins05} and the evolutionary models from \cite{schaller92}, then we can conclude that the cluster must be younger than $\sim$ 10 Myr. For older ages this O9V star would have evolved into a giant. This is consistent with the lack of red supergiants in the cluster, that need slightly less than 10 Myr to develop from 25 M$_\odot$ stars in the \citet{schaller92} models (including rotation in the evolutionary models does not change this conclusion).

\subsection{Mass}

We have estimated the stellar cluster mass of G61.48+0.09 assuming a Salpeter IMF. Further, we also assume that the mass
interval for O9V stars, the earliest type we have found, is 15.6-18.8 M$_\odot$ \citep[from the observational scale in][]{martins05}. 
With these assumptions the estimated mass contained in stars for G61.48+0.09 is 
1.5$\pm0.5 \times$10$^3$ M$_\odot$, where the error has been estimated from our adopted 
uncertainty of $\pm$1 spectral subtype. This error is larger than the one derived from changing 
the spectral type-stellar mass relationship, but other important sources may contribute to enlarge 
the uncertainty and make our estimation a lower limit.

The first one comes from the slope of the IMF. We have assumed a Salpeter form, which is in agreement
with most determinations for clusters in the mass range we investigate in G61.48+0.09 \citep{massey03}.
Evidences of anomalous, flatter IMFs are found only in more massive clusters \citep{figer05}. 
Therefore we consider the Salpeter form adequate for G61.48+0.09.

The second one comes from the possible presence of other early-type stars. We have seen in Section~\ref{cmd}
that the CMD stellar population above stars $\#$2 and $\#12$ is consistent with the nearby 
field population, which is also confirmed by the classification of star $\#8$ as a M2 III star. 
Therefore we don't expect a contribution of these stars to the early spectral type population of 
G61.48+0.09. More difficult is the possible contribution of star $\#$1. If the strong IR excess is
due to a disk around a young star, then its mass is probably not higher than 20 M$_\odot$ \citep{cesar06}. 
Assuming a second object of the same mass as star $\#12$ we would obtain a cluster mass
of 2900 M$_\odot$ (again with a Salpeter IMF). 

The final uncertainty would come from the lack of an evident stellar ionizing source. The ionizing
fluxes derived by \cite{puga04} cannot be fitted by  stars $\#1$ and $\#2$ (Puga et al.
stars 82 and 83) at the spectral types and luminosity classes we have assigned them. Therefore,
we have to look for another ionizing source, that presumably would be of earlier spectral type
and higher mass. \cite{puga04} propose that the source they identify in the L$^\prime$-band
as L1 could be the source. Inspection of the Spitzer archive does not help, as the region is
saturated at the long wavelengths. Therefore, we emphasize that the mass we derive for the cluster
is a lower limit. 

\section{Conclusions}
\label{conc}

A study of the stellar population of G61.48+0.09 has been done. This is part of our study of massive star
cluster candidates in the Milky Way that is being carried out using LIRIS at WHT. New near$-$IR photometric data 
and, for the first time, near$-$IR spectroscopy of stars in the cluster region have been presented. 

From the study of the Color-Magnitude Diagram and the spectroscopic analysis, we conclude that G61.48+0.09 is a star forming region 
located at 2.5$\pm$0.4 Kpc, in agreement with authors that have chosen the short distance scale \citep[see][]{deha00}. The
cluster is younger than 10 Myr and we obtain a stellar mass of 1.5$\times$10$^{3}$ M$_\odot$, although with large
uncertainty, as we cannot see the whole stellar content of G61.48+0.09. In particular, our classification of observed spectral types
cannot account for the ionizing fluxes derived by \cite{puga04} and it is unlikely that any of the stars observed in the near IR
can be the ionizing source. Table~\ref{summaryTable} lists a summary of all cluster properties.

\begin{acknowledgements}

We thank E. Puga for her careful reading of the manuscript and useful comments.
AMF would like to thank C. Warner, A. Gonzalez and the FLAMINGOS-2 team for providing access to the pre-release version of the FATBOY pipeline.
SR was supported by the MAEC-AECID scholarship. Part of this work was supported by the Science and Technology Ministry of the Kingdom of Spain (grants AYA2007-67456-C02-01/02 and AYA2008-06166-C03-01/02 and Consolider-Ingenio 2010 Program CSD 2006-00070 ) and by NASA under award NNG 05-GC37G, through the Long-Term Space Astrophysics program.
The William Herschel Telescope is operated on the island of La Palma by the Isaac Newton Group in the Spanish Observatorio del Roque de los Muchachos of the Instituto de Astrof\' \i sica de Canarias.

\end{acknowledgements}

\clearpage

\clearpage

\clearpage

\clearpage

\clearpage

\clearpage

\clearpage

\clearpage

\clearpage

\clearpage

\clearpage

\begin{table*}[h!]
\begin{minipage}[t]{\columnwidth}
\caption{Imaging/spectroscopic observations details. \label{TableObs}}                 
\centering                          
\begin{tabular}{|c c c | c | c | c | c | c |}        
Target      &  RA (J2000)  &  Dec (J2000)  &  Date   &   Filter/Grism   &    Exp. Time (s) & FWHM ($\arcsec$) & Standard star\\
\hline                        
G61.48+0.09    & 19 46 47   & +25 12 43  &   July 22, 2006    &     $J$ Filter           &  192.0  & 0.6 & $-$ \\
                           &                     &                     &   July 22, 2006     &    $H$ Filter            &  70.4  & 0.6 &  $-$\\
                           &                     &                     &   July 22, 2006     &    $K_S$ Filter      &  70.4 & 0.5 &  $-$\\
Control field  & 19 45 56 & +25 12 41       &    June 6, 2007    &    $J$ Filter            &  192.0  & 0.9 &  $-$\\
                           &                     &                     &   June 6, 2007     &    $H$ Filter            &  70.4  & 0.8 & $-$\\
                           &                     &                     &   June 6, 2007     &    $K_S$ Filter       &  70.4  &0.9  & $-$\\
\hline
G61.48+0.09    & 19 46 47   & +25 12 43  &   June 5, 2007     &     $HK$, R=945 (LS)       &            1200                   &    1.0     & HD229700\\
                           &                     &                     &   June 5, 2007     &     $K$, R=2500  (LS)     &            2400                   &       1.0   & HD229700\\
                           &                     &                     &   Sept. 20, 2007     &     $HK$, R=945  (MOS)     &           2400                    &   0.7  &   HD182761 \\
                           &                     &                     &   Sept. 20, 2007     &     $K$, R=2500   (MOS)    &            2400                   &     0.7   & HD182761 \\
\end{tabular}
\end{minipage}
\end{table*}

\clearpage

\begin{table*}[htdp]
\caption{Spectroscopically observed stars in G61.48+0.09. \label{spectab}}
\begin{center}
\begin{tabular}{|r|c|c|c|c|c|l|l|}
Star id & RA (2000) & $\delta$(2000) & $J$ & $H$ & $K_S$ & SpT & Comments \\
\hline
  1 & 19:46:47.585   & +25:12:45.64 & 13.50  & 12.06 &	 9.35 & YSO  & long-slit; $\#82$ in \cite{puga04} \\ 
14 & 19:46:47.823	& +25:12:29.98 & 13.94  & 12.34 & 11.85 & Be  & \\
12 & 19:46:41.550	& +25:12:20.52 & 14.60  & 12.25 & 10.92 & O9 V & \\
  2 &  19:46:47.311   & +25:12:45.60 & 12.60  & 11.65 &   9.47 & B0 V & long-slit; $\#83$ in \cite{puga04} \\
 4 & 19:46:48.993	& +25:13:28.86 & 13.97  & 12.46 &	11.53 &  B1 V & \\
  5 & 19:46:50.832   & +25:11:53.57 &  13.06   & 11.93	& 11.25	&  B2 V & \\
  3 & 19:46:47.033   & +25:12:45.55 & 12.58  & 12.52 & 10.40	& B3-B5 V  & long-slit \\ 
22 & 19:46:44.779	& +25:12:46.70 &  13.46   & 12.54 &	 12.11 & B3-B5 V & \\
17 & 19:46:50.608	& +25:13:11.86 &   15.85  & 13.93 & 12.73	& F5 III     & \\
  9 & 19:46:47.272	& +25:12:59.85 & 15.96    & 13.37 &	 11.77 & G3 III & \\
13 & 19:46:47.118	& +25:12:34.05 &  14.20   & 12.80 & 11.85 & G4 III & \\
  7 & 19:46:49.037	& +25:13:00.72 & 12.11   & 11.44 & 11.22	 & G5 III & \\
24 & 19:46:38.834	& +25:11:12.97 &  10.07   & 9.38 &	9.15 & G6 III & \\
18 & 19:46:51.209	& +25:14:11.36 & 17.83    & 13.76 &	 11.91 & G7 III & \\
25 & 19:46:40.728	& +25:10:57.91 &  15.39  & 13.24	& 12.32	 & G8 III & \\
27 & 19:46:46.924	& +25:12:13.37 &  10.40   & 9.64 &	9.47 & G9 III & \\
26 & 19:46:40.380	& +25:13:21.48 &  15.97   & 13.10 &	 11.68 & K0 III & \\
15 & 19:46:42.791	& +25:11:07.18 & 14.06    & 11.91 & 10.96	& K3 III & \\
  8 & 19:46:44.697	& +25:12:13.08 &  14.45   & 11.00	& 9.35	& M2 III & \\
  6 & 19:46:45.611	& +25:13:05.80 & 13.77	& 12.24 &	10.74 &  M3 III & \\
\end{tabular}
\end{center}
 \end{table*}%

\clearpage

\begin{table*}[htdp]
\caption{Derived extinctions and distances for the cluster stars. \label{ageTab}}
\begin{center}
\begin{tabular}{|r|c|c|}
Star id & $A_{K_S}$(mag) & d (Kpc) \\
\hline
12 & 2.6   & 2.1  \\
2   & 1.7  & 2.8  \\
4   & 1.6  & 3.1  \\
5   & 1.3  & 2.0  \\ 
3   & 1.5  & 2.1  \\
22 & 0.9  & 2.6  
\end{tabular}
\end{center}
 \end{table*}%

\clearpage

\begin{table*}[htdp]
\caption{Summary of cluster properties. \label{summaryTable}}
\begin{center}
\begin{tabular}{|r|c|}
Extinction & Varies from $A_{K_S}$ = 0.9 to 2.6 mag     \\
Distance   & 2.5$\pm$0.4 Kpc    \\
Age   & Younger than 10 Myr    \\
Minimum Mass   & 1.5$\times$10$^{3}$ M$_\odot$    
\end{tabular}
\end{center}
 \end{table*}%

\clearpage

\end{document}